\documentclass[aps,prd]{revtex4-2}

\usepackage{amsmath}  
\usepackage{amsfonts} 
\usepackage{graphicx} 
\usepackage[usenames,dvipsnames]{xcolor} 
\begin{document}


\title{Dual entropy gain by radiation at the zero temperature limit}

\author{Koray D\"{u}zta\c{s}}
\email{koray.duztas@okan.edu.tr} 
\affiliation{Faculty of Engineering and Natural Sciences, \.{I}stanbul Okan University, 34959 Tuzla \.{I}stanbul, T\"{u}rkiye}



\date{\today}

\begin{abstract}
Recently we have demonstrated that black hole radiation can be smoothly extended to the zero-temperature limit. Here we analyse entropy variations at the zero temperature limit and evaluate the validity of the Generalized Second Law (GSL). Typically one verifies that the entropy gain in the exterior region compensates for the entropy loss of the black hole due to evaporation. However an anomalous feature emerges at the zero temperature limit. The fact that the emission is restricted to the modes $\omega <m\Omega$, renders the variation of the horizon area of a Kerr black hole positive definite. Simultaneously the von Neumann entropy of the exterior region increases both for bosonic and fermionic modes. The dual entropy gain at the zero temperature limit marks a robust conservation of the GSL. We demonstrate that the dual increase  also applies to Kerr-Taub-NUT black holes whose entropy is neither directly proportional to the horizon area nor analytically integrable. 
\end{abstract}

\maketitle 
                  	  	                                                                                                                                                                                        
\section{Introduction}
Hawking’s area theorem which dictates that the total horizon area $A$ of a black hole must never decrease under processes satisfying the null energy condition \cite{hawkarea}, is reminiscent of the second law of thermodynamics if one establishes an analogy between the horizon area and entropy. Realizing the strength of this analogy,  Bekenstein postulated that black holes possess an inherent entropy proportional to their horizon area, heuristically estimating the proportionality constant as $(\ln 2)/(8 \pi)$ \cite{pagerev}. To prevent the violation of the standard second law when matter falls into a horizon, Bekenstein proposed the Generalized Second Law (GSL) \cite{beken1,beken2,beken3}, asserting that the joint entropy of the black hole ($S_{\rm{BH}}$) and its exterior surroundings ($S_{\rm{ext}}$) cannot decrease: 
\begin{equation}
\Delta (S_{\rm{BH}}+S_{\rm{ext}})\geq 0,
\label{beken1}
\end{equation}
Initially, the correspondence between the laws of mechanics and thermodynamics was viewed as a purely formal, mathematical analogy; a literal physical interpretation was resisted because classical black holes cannot radiate and thus possess an effective temperature of absolute zero \cite{thermo}. This perspective was transformed into a physical reality by Hawking’s discovery that quantum field effects during gravitational collapse cause black holes to emit thermal radiation \cite{hawkingorig}. 
\begin{equation}
N_{\omega lm}=\frac{\Gamma_{lm}(\omega)}{\exp [2\pi (\omega - m\Omega)/\kappa] \mp 1}.
\label{hawk1}
\end{equation}
This quantum emission confirmed a non-zero physical temperature proportional to the surface gravity, $T=\kappa /(2 \pi)$, precisely fixing the Bekenstein-Hawking entropy as:
\begin{equation}
S=\frac{A(kc^3)}{4(G \hbar)}.
\label{entropykerr}
\end{equation}
By defining entropy as a physical quantity rather than a geometric proxy, this formula unifies statistical mechanics, general relativity, and quantum mechanics, establishing a rigorous framework that continues to challenge our understanding of spacetime and information.

Although Hawking radiation appears to violate the classical area theorem \cite{hawkarea} by causing the black hole to shrink and lose entropy, the Generalized Second Law (GSL) remains valid due to a compensating entropy increase in the exterior region. The validity of the GSL during quantum evaporation can be verified by evaluating the von Neumann entropy of the thermal density matrix for each radiation mode \cite{pagerev}:
\begin{equation}
\delta S_{\rm{rad}} = -\sum P_n \ln P_n = (N \pm 1) \ln (1 \pm N) - N\ln N,
\label{vonneuman1}
\end{equation}
where $P_n$ denotes the probability of finding $n$ particles in a specific mode $(\omega,l,m)$, and the upper and lower signs correspond to bosonic and fermionic fields, respectively. Consequently, the entropy carried away by the emitted quanta offsets the reduction in the black hole's horizon area, ensuring the preservation of the GSL.

The GSL is also expected to hold in processes where black holes absorb test bodies or fields. In this case, the increase in the entropy of a black hole should compensate for the decrease in the entropy of the exterior region. To test the validity of the GSL, Bekenstein constructed a thought experiment where a test body is kept stationary near a black hole, then allowed to be absorbed \cite{bekenbound}. He proposed the existence of an upper bound on the entropy of the test body to preserve the validity of the GSL, which  turned out to be a controversial issue \cite{pagecomment,deutch,waldunruhbuo1,waldunruhbuo2,bekenbuo1,bekenbuo2}. Even in the absence of rigorous proofs, entropy bounds and the GSL has served as essential frameworks to explore the semi-classical and quantum nature of black holes \cite{gaowald,gsl1,gsl2,gsl3,gsl4,gsl5,boussorev,gsl6,gsl7,gsl8,gsl9,hodgsl}. ( See  \cite{wallten} for the critique of ten attempts to prove the GSL.) In a recent work, we evaluated the absorption of test fields by Kerr black holes to test the validity of the GSL \cite{duztasgsl}. We derived that the absorption of the fermionic fields with frequencies below the superradiance limit leads the area of the black hole to decrease. In this process, the black hole and the environment simultaneously lose entropy, which marks a generic violation of the GSL.  The derivation implies the validity of the GSL is restricted to the processes satisfying the null  energy condition, which can be circumvented by fermionic fields. The result also invalidates the proposals  to substitute the assumptions of the energy conditions by the GSL \cite{wallsing}. 

The expected number of particles emitted during Hawking radiation given in (\ref{hawk1}), becomes undefined at the point $\kappa=T=0$. However, we recently derived that the $T \to 0$ limit of the function $N(\omega,l,m)$ is well defined and radiation smoothly continues at the zero temperature limit \cite{hawkzero}. Consider the $T \to 0$ limit of the exponential function in the denominator in (\ref{hawk1})
\begin{equation}
\lim_{T \to 0} \exp \left( \frac{w-m\Omega}{T} \right) \left \{ \begin{array}{ll} \to \infty & (w-m\Omega>0) \\ =0 & (\omega -m\Omega <0) \end{array} \right. .
\end{equation}
The divergence of the exponential  for the modes $\omega >m\Omega$ implies that the average number of particles emitted $N$ in (\ref{hawk1}), converges to zero. Therefore the emission of the modes $\omega >m\Omega$ is suppressed at the zero temperature limit. We proceed with the modes $\omega <m\Omega$. Though the exponential is undefined at $T=0$ --which corresponds to extremal black holes--, the limit is is well defined. One derives
\begin{eqnarray}
&& \lim_{T \to 0} N^B_{\omega lm}=\frac{\Gamma_{lm}}{-1}= \vert \Gamma_{lm} \vert \nonumber \\
&&\lim_{T \to 0}N^F_{\omega lm}=\frac{\Gamma_{lm}}{1}= \Gamma_{lm},
\label{nzero}
\end{eqnarray}
where the superscripts $B$ and $F$ refer to bosons and fermions, respectively. The emission at the $T \to 0$ limit is restricted to the modes with $\omega < m\Omega$, for which the exponential does not diverge. Note that the absorption probability $\Gamma$ becomes negative for bosonic modes in the range $\omega< m\Omega$ while it remains positive for fermionic modes, which ensures that the expected number of emitted particles is positive in both cases.

Here, we evaluate the entropy variations of Kerr and Kerr-Taub-NUT black holes during the counter-intuitive emission at the zero temperature limit. We would like to note that the analysis applies to the extremal limit $T \to 0$, rather than extremal black holes with $T=0$. Though, the temperature is not exactly zero, it is vanishingly small relative to any arbitrary power of a small positive parameter, $T \ll \varepsilon^n $. (Here we let $M=1$ so that the temperature is dimensionless.) In this limit, the first law of black hole mechanics becomes ill-defined, as made explicit when written in the form:
\begin{equation}
\delta S_{\rm{BH}}=\frac{1}{T}(\delta M - \Omega \delta J).
\label{firstlaw}
\end{equation}
Though $\delta M$ is a first order quantity in the test particle/field approximation, $T \ll \delta M$ at the zero temperature limit. Therefore the first law implies an unphysical divergence in the entropy of the black hole at the zero temperature limit, which renders it ill-defined. Here, we directly evaluate the area variation to calculate $\delta S_{\rm{BH}}$, as one cannot employ the first law at the zero temperature limit. 

In section (\ref{kerrzero}), we evaluate the variations in the area/entropy of a Kerr black hole and carry out an explicit analysis for the most probable mode. In section (\ref{ktn}) we extend the analysis to Kerr-Taub-NUT black holes whose entropy is not directly proportional to the horizon area.

\section{Radiation at the $T \to 0$ limit and entropy}
\label{kerrzero} 
In this section, we analyse the variation in the entropy of a Kerr black hole at the zero temperature limit, during the emission of bosonic and fermionic modes. The entropy of a Kerr black hole is given by  (\ref{entropykerr}), which reduces to $S=A/4 $ when one adapts the natural units $G=\hbar=c=1$. Recently we evaluated the area/entropy variation of a Kerr black hole as it absorbs bosonic and fermionic test fields \cite{duztasgsl}. However the radiation is emitted in the current analysis, which implies that the variations in the mass ($\delta M$) and the angular momentum ($\delta J$) parameters are negative. The variations are related by
\begin{equation}
\delta J=\frac{m}{\omega}\delta M,
\label{deltajdeltam}
\end{equation}
where $\omega$ is the frequency of the emitted mode, and $m$ is the azimuthal quantum number. We start with the horizon area of a Kerr black hole 
\begin{equation}
A=4\pi(r_+^2 +a^2)=8\pi (M^2+\sqrt{M^4-J^2}),
\end{equation}
where $r_+=M+\sqrt{M^2-a^2}$ is the radial coordinate of the event horizon, and $a=J/M$ is the Kerr parameter. We consider the variation of the area with respect to the mass and angular momentum parameters. Imposing (\ref{deltajdeltam}), one derives \cite{duztasgsl}
\begin{equation}
\delta A=8 \pi \frac{\left[2M (M+\sqrt{M^2-a^2}) -a\left(\frac{m}{\omega}\right)\right] }{\sqrt{M^2-a^2}} \delta M .
\label{deltaA1}
\end{equation}
Note that $\delta A$ vanishes at $\omega = m\Omega=(ma)/(2Mr_+)$. Considering the fact that  $\delta M$ is negative during emission, the variation in the area becomes positive for the modes $\omega < m\Omega$, and negative for the modes $\omega > m\Omega$. For any non-zero $T$,  the average number of particles emitted in the modes  $\omega > m\Omega$ is positive-definite for both bosonic and fermionic fields. Therefore the area can increase or decrease depending on the frequency of the emitted mode. However, the emission of the modes  $\omega > m\Omega$ is suppressed at the zero temperature limit. The fact that the frequencies of the emitted mode are restricted to $\omega < m\Omega$ at the zero temperature limit implies that the variation in the area is positive-definite 
\begin{equation}
\lim_{T \to 0} \delta A > 0 .
\label{limitzerokerr}
\end{equation}
The positive-definiteness of $\delta A$ --or equivalently $\delta S_{BH}$-- is a distinct feature of the zero-temperature limit, which could be violated for any finite positive temperature by the emission of the modes $\omega > m\Omega$. Therefore it should not be interpreted as an analogue of the classical area theorem for Hawking radiation. Note that the expected number of particles at the zero temperature given in (\ref{nzero}) satisfies $0<N<1$. One can substitute any number in the range $\{ 0,1\}$ for $N$ in (\ref{vonneuman1}) , and verify that the von-Neumann entropy is positive for both bosonic and fermionic fields. Therefore the entropy of the exterior region increases. Meanwhile the entropy of the black hole also increases due to the emission of the modes with $\omega <m \Omega$. This simultaneous increase at the zero temperature limit marks a robust conservation of the GSL.

\subsection{Explicit analysis for $\omega=m\Omega /2$}
In our recent study on Hawking radiation at the zero temperature limit, we derived that the highest probability of emission pertains to the modes $\omega=m\Omega /2$, both for bosonic and fermionic cases. Here, we carry out an explicit calculation for the area/entropy variation during the emission of these modes. We start with a nearly extremal Kerr black hole parametrized as
\begin{equation}
M^2-\frac{J^2}{M^2}=M^2 \epsilon^2,
\label{param1}
\end{equation}
where the limit $\epsilon \to 0$ corresponds to the extremal limit of zero temperature. The initial area of the black hole is given by
\begin{equation}
A_{\rm{in}}=8\pi M^2 (1+\epsilon)
\end{equation}
which reduces to $A=8\pi M^2$ at the zero temperature limit. The emitted particles carry energy $\delta M$ and angular momentum $\delta J$. The final parameters of the space-time satisfy
\begin{eqnarray}
M_{\rm{fin}} &=& M-\delta M -\delta^2 M, \nonumber \\
M_{\rm{fin}}^2&=&M^2+ (\delta M)^2 -2M\delta M -2M \delta^2 M, \nonumber \\
J_{\rm{fin}} &=& J- \delta J- \delta^2 J, \nonumber \\
J_{\rm{fin}}^2&=&J^2+ (\delta J)^2 -2J\delta J -2J \delta^2 J,
\label{mfinjfin}
\end{eqnarray}
The horizon area after the emission is given by:
\begin{equation}
A_{\rm{fin}}=8 \pi \left( M_{\rm{fin}}^2 + M_{\rm{fin}} \sqrt{M_{\rm{fin}}^2 - \frac{J_{\rm{fin}}^2}{M_{\rm{fin}}^2}} \right)
\label{afinal}
\end{equation}
We parametrize the contributions of the test field to the mass and angular momentum parameters as follows:
\begin{equation}
\delta M=M\eta, \quad \delta J=\frac{m}{\omega}\delta M,
\label{paramfield1}
\end{equation}
where $\eta \ll 1$ in accord with the test field approximation. Note that we use different parameters for the closeness to extremality ($\epsilon$) and the contributions of the test field ($\eta$). The zero temperature limit  is identified by $\epsilon \to 0$, without referring to $\eta$. First we evaluate the argument of the square root in (\ref{afinal}). Note that to second order:
\begin{equation}
\frac{J_{\rm{fin}}^2}{M_{\rm{fin}}^2}=\frac{J_{\rm{fin}}^2}{M^2}\left( 1+2\eta +3\eta^2 +2\frac{\delta^2 M}{M} \right),
\end{equation}
where $J_{\rm{fin}}^2$ is given in (\ref{mfinjfin}). We use the parametrization (\ref{paramfield1}) and impose that $\omega=m\Omega /2$ for the most probable mode of emission. Note that $\Omega=m/2M$ in the extremal limit, leading to
\begin{equation}
\delta J=\frac{m}{\omega}\delta M=4M^2 \eta
\label{deltaj}
\end{equation}
we also substitute $J^2=M^4(1-\epsilon^2)$ and $J=M^2(1-\epsilon^2 /2)$, which directly follow from the parametrization (\ref{param1}). After some algebra, one derives
\begin{eqnarray}
M_{\rm{fin}}^2 -\frac{J_{\rm{fin}}^2}{M_{\rm{fin}}^2}&=& M^2\epsilon^2  +4M^2\eta -2M^2\eta^2 \nonumber  \\ 
&-&4M\delta^2 M +2 \delta^2 J .
\label{mfinjfin2}
\end{eqnarray}
The expression (\ref{mfinjfin2}) involves second order perturbations. In \cite{spin2} we evaluated the second order perturbations in Kerr space-time to derive that
\begin{equation}
2M\delta^2 M-\delta^2 J=\frac{(\delta J)^2}{2M^2},
\end{equation}
which implies
\begin{equation}
4M\delta^2 M -2 \delta^2 J=16M^2\eta^2,
\end{equation}
for $\delta M=M\eta$ and $\delta J$ given by (\ref{deltaj}). We take the limit $\epsilon \to 0$ and  -evaluate the argument of the square root. We derive that
\begin{equation}
\sqrt{M_{\rm{fin}}^2 - \frac{J_{\rm{fin}}^2}{M_{\rm{fin}}^2}}=2M \left( \sqrt{\eta} - \frac{9}{4}\eta^{3/2} \right),
\end{equation}
where we used the approximation
\[
\sqrt{\eta - \frac{18}{4}\eta^2} \simeq \sqrt{\eta} - \frac{9}{4}\eta^{3/2}.
\]
We proceed to calculate $A_{\rm{fin}}$. Note that
\begin{eqnarray}
&& M_{\rm{fin}}^2 + M_{\rm{fin}} \sqrt{M_{\rm{fin}}^2 - \frac{J_{\rm{fin}}^2}{M_{\rm{fin}}^2}}=M^2 +M^2\eta^2 -2M^2 \eta -2M \delta^2 M \nonumber \\
&+&(M+M\eta +\delta^2 M)\left( \sqrt{\eta} -\frac{9}{2}M \eta^{3/2} \right).
\end{eqnarray}
To first order, this implies
\begin{equation}
\lim_{\epsilon \to 0} A_{\rm{fin}}=8 \pi M^2 \left( 1+2\sqrt{\eta} - 2\eta \right).
\end{equation}
Considering the fact that $\sqrt{\eta} > \eta $ for $\eta \ll 1$, one concludes that
\begin{equation}
 A_{\rm{fin}}> A_{\rm{in}}.
\end{equation}
The area/entropy of the black hole increases as a result of the emission of the modes with frequency $\omega=m\Omega /2$. Simultaneously the von Neumann entropy (\ref{vonneuman1}) of the exterior region increases, marking a robust conservation of the GSL.
\section{Kerr-Taub-NUT black holes}
\label{ktn}
Kerr-Taub-NUT (KTN) space-time is a type D vacuum solution of Einstein's field equation endowed with a gravito-magnetic charge (NUT charge) $\ell$ \cite{demianski}. The space-time does not encapsulate a curvature singularity provided that $a<\vert \ell \vert$, where $a=J/M$ is the Kerr parameter. However the space-time involves incomplete null geodesics imprisoned in a compact neighbourhood of the event horizon whose affine parameter cannot be extended beyond a finite value, which renders it singular according to the definition of Penrose and Hawking. (See e.g. \cite{hawkellis}) The entropy of a KTN black hole is neither directly proportional to its horizon area, nor analytically integrable. The entropy variation can be expressed as \cite{frodden}
\begin{eqnarray}
\delta S&=&\frac{2 \pi}{\sqrt{M^2 -a^2 +\ell^2}}\nonumber \\
&\times&\left[(M+\sqrt{M^2 -a^2 +\ell^2}-a^2 +2\ell^2)2M\delta M - ma\delta a \right ] \nonumber \\
& &
\end{eqnarray}
By direct substitution $a=J/M$ and $\delta J =(m/\omega) \delta M$, one derives
\begin{eqnarray}
\delta S&=&\frac{2 \pi}{\sqrt{M^2 -a^2 +\ell^2}} \nonumber \\
&\times &  \left[2M(M+\sqrt{M^2 -a^2 +\ell^2})+2\ell^2-\frac{ma}{\omega} \right]\delta M. \nonumber \\
\label{deltasktn}
\end{eqnarray}
Note that the angular velocity and the spatial coordinate of the event horizon are
\begin{equation}
\Omega= \frac{ma}{2(Mr_+ + \ell^2)}, \quad r_+=M+\sqrt{M^2 +\ell^2 -a^2},
\end{equation}
respectively. Considering the fact that $\delta M$ is negative in the case of emission, (\ref{deltasktn}) implies
\begin{equation}
\omega < m\Omega \rightarrow \delta S > 0.
\end{equation}
The entropy of the KTN black hole increases if the emitted modes satisfy $\omega < m \Omega$, which is peculiar to the radiation at the zero temperature limit. Therefore simultaneous increase in the entropy of the black hole and the exterior region also applies to KTN space-time whose entropy is not directly proportional to its horizon area.
\section{Conclusions}
Previously, we have demonstrated that Hawking radiation can be extended to the zero temperature limit \cite{hawkzero}. In this work we analysed the entropy variation during the evaporation at the zero temperature limit where the emission is restricted to the modes $\omega < m\Omega$. In similar attempts to evaluate the validity of the GSL, the radiation naturally induces a decrease in the area/entropy of the black hole and one checks if the entropy gain of the exterior region can compensate for this decrease, to re-assure the validity of the GSL. However, we showed that the fact that the emission is restricted to the modes $\omega < m\Omega$ at the zero temperature limit leads the area/entropy of a Kerr black hole to increase. Simultaneously the von Neumann entropy of the exterior region also increases both for fermionic and bosonic modes. The dual entropy gain at the zero temperature limit marks a robust conservation of the GSL for Kerr black holes. We carried out an explicit analysis for the modes with $\omega=m\Omega /2$ which possess the highest probability of emission to justify our claim. We extended the analysis to KTN black holes whose entropy is neither directly proportional to the area, nor analytically integrable. We demonstrated that the dual entropy gain at the zero temperature limit applies to KTN space-time as well.

\textbf{Data Availability Statement} This manuscript has no associated data. [Author's comment: This manuscript has no associated data since it is purely theoretical.]

\textbf{Code Availability Statement} This manuscript has no associated code/software. [Author’s comment: No code/software has been generated or analyzed during the current study.]


\begin{thebibliography}{99}

\bibitem{hawkarea}  S.W. Hawking, Phys. Rev. Lett. \textbf{26} 1971, 1344 (1971).

\bibitem{pagerev} D.N. Page, New J. Phys. \textbf{7}, 203 (2005)

\bibitem{beken1}J.D. Bekenstein, Lett. Nuovo Cimento \textbf{4}, 737 (1972).

\bibitem{beken2} J.D. Bekenstein, Phys. Rev. D \textbf{7}, 2333 (1973)

\bibitem{beken3} J.D. Bekenstein, Phys. Rev. D \textbf{9}, 3292 (1974)

\bibitem{thermo}J. Bardeen, B. Carter,  S.W. Hawking,   Comm. Math. Phys.    \textbf{31}, 161 (1973).

\bibitem{hawkingorig} S.W. Hawking, Commun. Math. Phys. \textbf{43}, 199 (1975).

\bibitem{bekenbound} J.D. Bekenstein, Phys. Rev. D \textbf{23}, 287 (1981).

\bibitem{pagecomment} D.N. Page, Phys. Rev. D \textbf{26}, 947 (1982).


\bibitem{deutch} D. Deutch, Phys. Rev. Lett. \textbf{48}, 286 (1982).

\bibitem{waldunruhbuo1} W.G. Unruh, R.M. Wald, Phys. Rev. D \textbf{25}, 942 (1982).

\bibitem{waldunruhbuo2} W.G. Unruh, R.M. Wald, Phys. Rev. D \textbf{27}, 2271 (1983).

\bibitem{bekenbuo1} J.D. Bekenstein, Phys. Rev. D \textbf{49}, 1912 (1994).

\bibitem{bekenbuo2} J.D. Bekenstein, Phys. Rev. D \textbf{60}, 124010 (1999).

\bibitem{gaowald} S. Gao, R.M. Wald, Phys. Rev. D \textbf{64}, 084020 (2001).

\bibitem{gsl1} C. Eling, J.D. Bekenstein, Phys. Rev. D \textbf{79},  024019 (2009).

\bibitem{gsl2} R.	Bousso, N. Engelhardt,  Phys. Rev. D \textbf{93},  024025 (2016).

\bibitem{gsl3} L.H.	Ford, T.A. Roman, Phys. Rev. D \textbf{64},  024023 (2001).

\bibitem{gsl4}	J. Ho, Phys. Rev. D \textbf{64},  064019 (2001). 

\bibitem{gsl5} A.	Hosoya,  A. Carlini,  T. Shimomura,  Phys. Rev. D \textbf{63},  104008 (2001).  

\bibitem{boussorev} R. Bousso, Rev. Mod. Phys. \textbf{74}, 825 (2002).

\bibitem{gsl6} 	E.N. Saridakis,  S. Basilakos, Eur. Phys. J. C \textbf{81}, 644 (2021).

\bibitem{gsl7} C.	Tsallis,  L.J.L. Cirto,  Eur. Phys. J. C \textbf{73}, 2487 (2013). 

\bibitem{gsl8} D. Chen, H. Yang,  Eur. Phys. J. C \textbf{72}, 2027 (2012).  

\bibitem{gsl9} 	T. Tanaka, T. Tamaki, Eur. Phys. J. C \textbf{73}, 2314 (2013). 

\bibitem{hodgsl} S. Hod, Phys. Lett. B \textbf{751}, 241 (2015).

\bibitem{wallten} A.C. Wall, J. High energy Phys. \textbf{06}, 021 (2009).

\bibitem{duztasgsl} K. D\"{u}zta\c{s}, Eur. Phys. J. C \textbf{85}, 1073 (2025).

\bibitem{wallsing} A.C. Wall,  Class. Quantum Grav. \textbf{30}, 165003 (2013)

\bibitem{hawkzero} K. D\"{u}zta\c{s}, Eur. Phys. J. C \textbf{84}, 1164 (2024).

\bibitem{spin2} K. D\"{u}zta\c{s}, 	 Phys. Rev. D  \textbf{110}, 024081 (2024).

\bibitem{demianski} M. Demianski, E.T. Newman, Bull. Acad. Pol. Sci. Ser. Sci. Math. Astron. Phys. \textbf{14}, 653 (1966).

\bibitem{hawkellis} S.W. Hawking, G.F.R. Ellis, {\it The large scale structure of space-time}, Cambridge University Press, Cambridge, (1973).

\bibitem{frodden} E. Frodden, D. Hidalgo, Phys. Lett. B \textbf{832}, 137264 (2022).












\end{thebibliography}
\end{document}